\documentclass{article}
\pdfoutput=1

\usepackage{arxiv}
\usepackage{amsmath} 
\usepackage[utf8]{inputenc} 
\usepackage[T1]{fontenc}    
\usepackage{hyperref}       
\usepackage{url}            
\usepackage{booktabs}       
\usepackage{amsfonts}       
\usepackage{nicefrac}       
\usepackage{microtype}      
\usepackage{lipsum}
\usepackage{graphicx}
\usepackage{algorithm}%
\usepackage{algorithmicx}%
\usepackage{algpseudocode}%
\usepackage{listings}%
\usepackage{chemformula}
\usepackage{mhchem}
\usepackage{booktabs}%
\graphicspath{ {./images/} }

\title{What it takes to break a liquid: \\analysis of the cavitation threshold in various media }

\author{
 Gianmaria Viciconte \\
  Physical Science and Engineering, \\
  King Abdullah University of Science and Technology, \\ 
  Thuwal, 23955, Saudi Arabia \\
  \texttt{gianmaria.viciconte@kaust.edu.sa} \\
  \And
   Paolo Guida \\
  Physical Science and Engineering, \\
  King Abdullah University of Science and Technology, \\ 
  Thuwal, 23955, Saudi Arabia \\
  \texttt{paolo.guida@kaust.edu.sa} \\
   \And
   Tadd T. Truscott \\
  Physical Science and Engineering, \\
  King Abdullah University of Science and Technology, \\ 
  Thuwal, 23955, Saudi Arabia \\
  \texttt{tadd.truscott@kaust.edu.sa} \\
  \And
 William L. Roberts \\
  Physical Science and Engineering, \\
  King Abdullah University of Science and Technology, \\ 
  Thuwal, 23955, Saudi Arabia \\
  \texttt{william.roberts@kaust.edu.sa} \\
}

\begin{document}
\maketitle
\begin{abstract}
Cavitation has historically been related to parameters measured at equilibrium, such as vapor pressure and surface tension.  However, nucleation might occur when the liquid is metastable, especially for fast phenomena such as cavitation induced by high-frequency acoustic waves.
This is one of the reasons for the large discrepancy between the experimental estimate of the cavitation threshold and the theory's predictions.

Our investigation aims to identify nucleation thresholds in various substances characterized by different physical properties. The experiments were performed by initiating nucleation through ultrasound at 24 kHz. The cavitation onset was studied using a novel procedure based on high-speed imaging and acoustic measurements with a hydrophone. Combining these two techniques allowed us to define the exact instant cavitation occurred in the liquid medium. The bubble nucleation was framed at 200,000 fps with a spatial resolution in the order of micrometers. Such fine temporal and spatial resolutions allowed us to track the expansion of the cavitation bubble right after its onset. We tested five different substances and tracked the amplitude of the transducer oscillation to reconstruct the pressure field when cavitation occurs. This allows us to identify the liquid's acoustic cavitation threshold (tensile strength).
The data collected confirmed that the vapor pressure is not a good indicator of the occurrence of cavitation for acoustic systems. Furthermore, all substances exhibit similar behavior despite their different physical properties. This might seem counterintuitive, but it sheds light on the nucleation mechanism that originates cavitation in a lab-scale acoustic system.
\end{abstract}
\section{Introduction}\label{sec1}

Cavitation, traditionally studied in the field of hydrodynamics \cite{brennen2011hydrodynamics}, also occurs in liquids subjected to acoustic or pressure waves. When the pressure descends below a certain value, vapour-filled cavities form within the liquid \cite{brennen2014cavitation, urick1975principles}. These cavities oscillate in response to the periodic pressure field, storing acoustic energy that is locally released upon collapse. This release generates “hot spots” with extreme temperature and pressure, facilitating the formation of radical species that drive chemical reactions, a phenomenon known as sonochemistry \cite{suslick1990sonochemistry, suslick2008inside}. Additionally, acoustic cavitation produces unique fluid dynamics effects, such as micro-jets and shock waves \cite{wagterveld2011visualization}, which play a key role in particle fragmentation, enhanced chemical leaching, and various industrial processes \cite{suslick2008inside,dhanasekaran2024mechanistic, lithiumLeaching}.

Despite its broad industrial applications, the design and operation of ultrasonic reactors often rely on empirical methods. This limitation stems from an incomplete understanding of the underlying physical mechanisms involved in pressure-induced nucleation. In hydrodynamic cavitation, the threshold is commonly assumed to be equal to the saturation vapor pressure of the liquid \cite{brennen2011hydrodynamics}. However, this assumption does not hold for cavitation induced by acoustic waves, especially at high frequencies \cite{brennen2014cavitation, urick1975principles}. Liquids enter metastable states at such frequencies and resist phase change at pressures well below their saturation vapor pressure \cite{brennen2014cavitation, urick1975principles}. This resistance arises from the elastic behavior of liquids under high-frequency perturbations \cite{brennen2014cavitation}, as demonstrated by Urick \cite{urick1975principles}, who showed that the cavitation threshold significantly increases with wave frequency.

Experimental measurements of the cavitation threshold further underscore this complexity. Urick \cite{urick1975principles} reported values significantly lower than those observed in highly purified liquids. Experiments by Berthelot \cite{berthelot1850quelques} and Dixon \cite{dixon1909note} estimated thresholds of $-5$~MPa and $-20$~MPa, respectively, for purified water. More recent investigations by Caupin \textit{et al.} \cite{caupin2012exploring} confirmed thresholds near $-20$~MPa. These values, although obtained under controlled conditions, are substantially below theoretical predictions from intermolecular force models or Classical Nucleation Theory (CNT), which estimate thresholds as high as $100$~MPa \cite{brennen2014cavitation, skripov1974metastable, herbert2006cavitation}. 

While the vapor pressure does not coincide with the cavitation threshold, macroscopic thermophysical properties, particularly surface tension, are expected to play a more relevant role. According to the CNT, the cavitation threshold scales with surface tension to the power of 3/2 \cite{herbert2006cavitation, brennen2014cavitation, blander1979bubble, skripov1974metastable}. For heterogeneous nucleation, the Blake threshold indicates a linear relationship between surface tension and cavitation threshold \cite{louisnard2003growth, neppiras1980acoustic, akhatov1997role}. Recent studies suggest that models incorporating the Tolman length, which accounts for curvature effects on surface tension at the nanoscale, offer improved predictions \cite{kashchiev2020nucleation}. Molecular dynamics simulations further refine CNT, yielding higher nucleation rates at given conditions \cite{menzl2016molecular}.

The results detailed in this work explore how thermophysical properties affect the cavitation threshold in various substances. For this, we conducted experiments using a 24 \text{kHz} lab-scale horn-type ultrasonic device. The cavitation onset was identified using a novel methodology combining high-speed imaging and hydrophone measurements \cite{viciconte2025general,viciconte2025high}. The optical setup provided high temporal and spatial resolutions, enabling precise measurements of displacement amplitude (micrometer scale) and ultrasound probe frequency. The acoustic pressure at cavitation onset was calculated using the Rayleigh Integral solution \cite{rayleigh2024theory, sherman2007transducers, viciconte2023towards}. We tested various liquids with differing viscosities, including water, ethanol, acetone, isopropanol, dimethyl sulfoxide (DMSO), and silicon oils.

\begin{figure}
    \centering
    \includegraphics[width=0.6\linewidth]{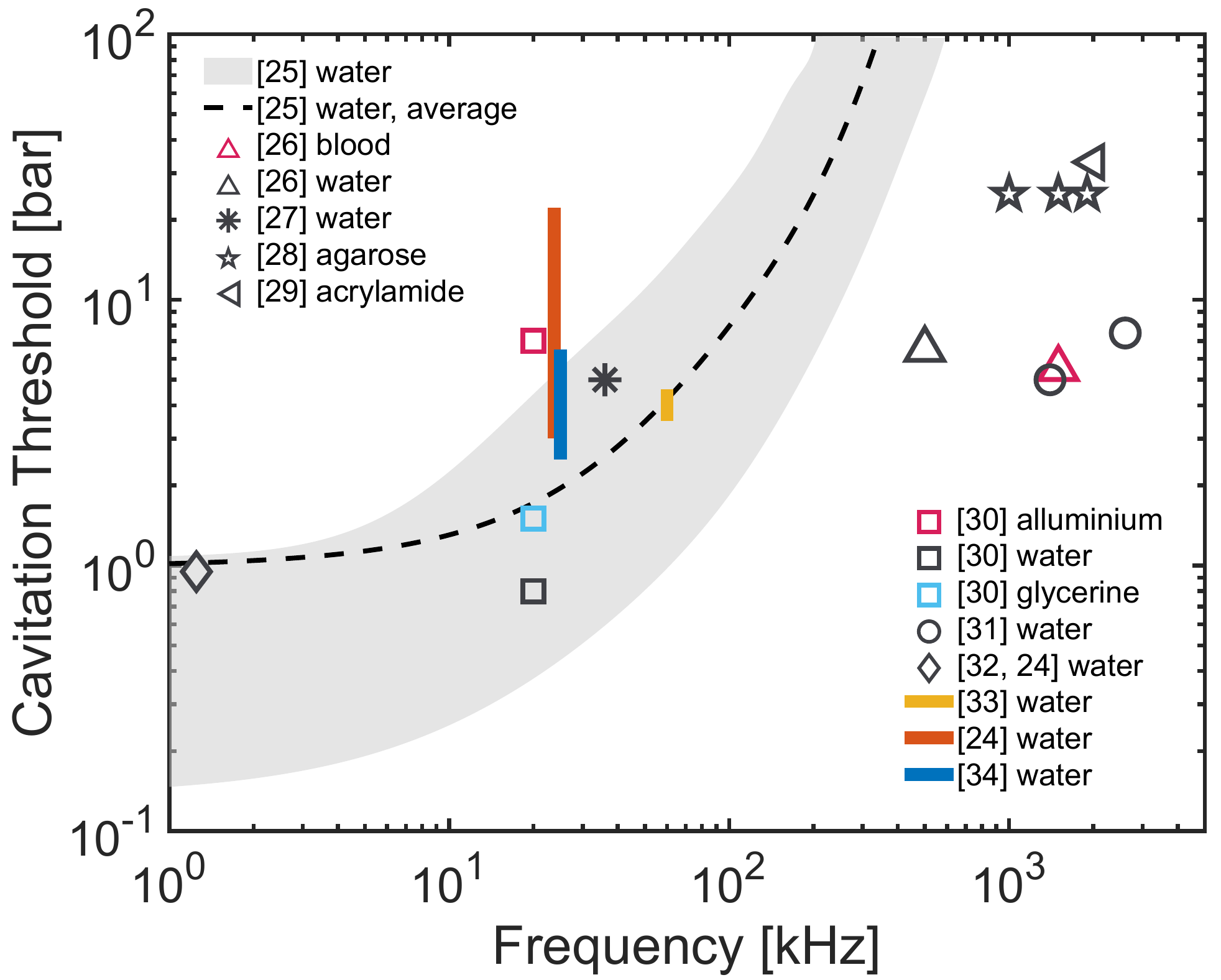}
\caption{Cavitation threshold measured for various substances as a function of the frequency. The data obtained by several authors are represented \cite{esche1952untersuchung, huang2022dynamic, crum1980acoustic, bawiec2021inertial, thomas2005dynamics, tzanakis2017characterizing, dong2020simulation, pan2017cavitation, viciconte2023towards, blake1949onset, strasberg1959onset}.}
\label{figure_1}
\end{figure}

Interestingly, despite variations in thermophysical properties, the tested liquids exhibited similar resistance to phase change. A literature survey confirmed what we observed as it is possible to appreciate in \ref{figure_1}, where we report the cavitation threshold measured by various authors for different substances and frequencies. The experimental evidence suggests that the cavitation threshold depends more on the perturbation rate than the fluid's intrinsic properties. Furthermore, solutions of the Rayleigh–Plesset indicate that liquids can endure greater pressure differentials at higher frequencies, underscoring the importance of frequency in determining cavitation behavior. In the next few sections, we try to explain the occurrence of such evidence.

\section{Materials and Methods}
\label{materialsAndMethods}
\subsection{Experimental setup}
The experimental campaign was conducted with the same backlighting setup previously used by the authors \cite{viciconte2023towards}. A Hielscher device UP400S (nominal frequency of 24 kHz) was used to generate ultrasound in the liquid domain. This device provides a maximum power of 400 W and can be equipped with titanium probes with different diameters. For the present study, a 3 mm cylindrical titanium probe was used. The arithmetical mean roughness (\(Ra\), Stand. JIS2001) of the emitting surface of the probe was characterized with the surface roughness tester Mitutoyo Surftest SJ-400, and it turned out to be equal to \(Ra\) = 0.14 $\mu$m. The liquid was placed in a Hellma transparent container 50$\times$50$\times$50 mm$^3$, made of optical glass. The ultrasonic probe was submerged 20 mm in the liquid. A LED light source (Godox SL-200W II), emitting white light non-coherently, was placed on the back side, aligned with the axis of the probe. The scene was captured with the high-speed camera Photron FASTCAM Nova S16 at 200,000 fps (128x224 pixels). The experiments have been conducted using substances with an high-level of purity: ethanol (absolute, $\geq$ 99.8\%, AnalaR NORMAPUR ACS, VWR chemicals), acetone (Honeywell for HPLC, $\geq$ 99.8\%), isopropanol (2-propanol, $\geq$ 99.8\%, HiPerSolv CHROMANORM for HPLC, VWR chemicals), and DMSO (dimethyl sulfoxide, puriss. p.a., dried, $\leq$ 0.02\% water, Sigma-Aldrich). Three different types of high-purity water were tested: Milli-Q water (total organic carbon (TOC) $\leq$ 5 ppb), water extra pure deionized Thermo Fisher (residue after evaporation $\leq$ 25 ppm), and pure water demineralized Thermo Fisher (Silicon dioxide (SiO2) $\leq$ 1.0 ppm). Furthermore, to evaluate the influence of dynamic viscosity on the cavitation threshold, silicon oils characterized by three different dynamic viscosities were tested: 5 cP (IKA CAL-O-5), 10 cP (IKA CAL-O-10), and 50 cP (Fungilab Viscosity Reference Standard, RT50).
The liquids were tested at a temperature of 24$^{\circ}$ C and atmospheric pressure. The dissolved air content of the liquids was not altered with preliminary treatments, and all the experiments were conducted at the saturation level.

\subsection{Estimation of the cavitation threshold}

\begin{figure*}[h!]
    \centering
\includegraphics[width=\linewidth]{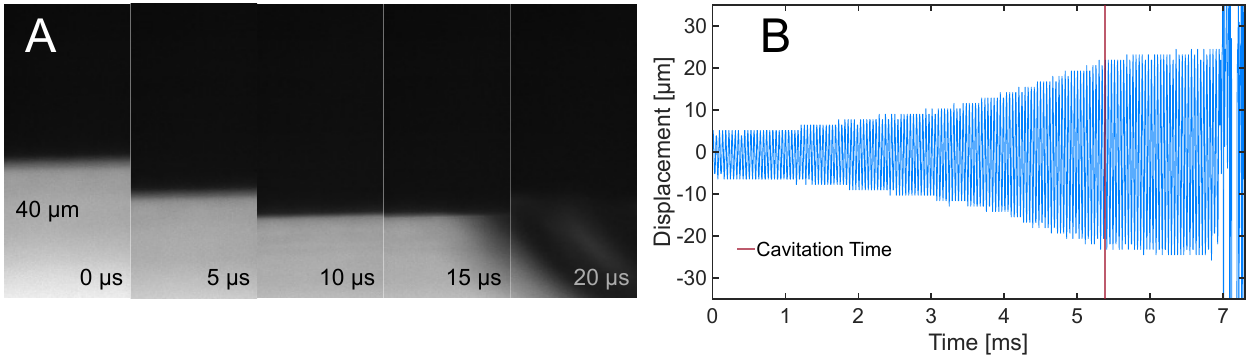}
    \caption{(B) Sequence of frames recorded by the high-speed camera, equipped with a 10$\times$ objective lens. The sequence is related to a 3 mm titanium probe immersed in isopropanol. (A) Result of the tracking procedure on the frames acquired with the high-speed camera. The blue function represents the displacement of the probe over time, while the vertical red line represents the time at which cavitation happens.}
    \label{figure2}
\end{figure*}

When the ultrasound device is turned on, the displacement amplitude \(A\) of the ultrasonic probe undergoes a transient state where it gradually increases until it reaches a steady state. For the device Hielscher UP400S, the nucleation of the first cavity always happens during the transitory regime. To estimate the acoustic pressure at which cavitation onset happens, the probe's displacement must be precisely measured. Since the displacement is in the order of magnitude of micrometers, the high-speed camera was equipped with a Mitutoyo 10X objective lens with a resolution of 1.74 $\mu$m/pixel. The sequence of frames, obtained for one experimental run, is shown in Fig. \ref{figure2}a. With such fine spatial and temporal resolutions, the probe tip's movement and the first cavity's nucleation, attached to the surface, were captured \cite{viciconte2023towards}.

The probe displacement was measured with a tracking procedure implemented on Matlab. Starting from the sequence of frames (Fig. \ref{figure2}a), the displacement can be tracked by selecting one column of pixels and by taking, along this column, the difference between adjacent pixels. The point at
which the difference is maximum defines the maximum gradient of pixel intensity and, therefore, the location of the probe
tip at every time step. This procedure can track the probe displacement over time until cavitation happens (Fig. \ref{figure2}b). The Discrete Fourier Transform (DFT) of the data yields a probe frequency of $f_p=23.78$ kHz. The red vertical line at 5.2 ms, in Fig. \ref{figure2}b, indicates the time of cavitation inception ($t_{cav}$), i.e., the time at which the first bubble appears in the frame sequence (Fig. \ref{figure2}a).

Once the cavitation time defined, the cavitation amplitude (\(A_{cav}\)) can be found by considering the nearest relative maxima and minima, in a neighborhood of \(t_{cav}\) (red line in Fig. \ref{figure2}b), and by averaging their absolute values.

The acoustic emission from a cylindrical probe can be modeled by considering its analogy with a plane circular transducer \cite{sherman2007transducers}, where the acoustic pressure field, on the probe axis, can be estimated by analytically solving the Rayleigh Integral \cite{sherman2007transducers, rayleigh2024theory, viciconte2023towards}. For the configuration adopted in the present study, the maximum acoustic pressure amplitude occurs on the probe surface, corresponding with the axis. Starting from the vibration frequency of the probe (\(f\)) and the displacement amplitude (\(A_{cav}\)), the acoustic pressure amplitude, that has generated cavitation (\(p_{0}|_{cav}\)), can be found with the following expression:
\begin{equation}
p_{0}|_{cav}= 2 \rho c \omega A_{cav}  \left|\sin\left(\frac{kd}{2}\right)\right|,
\label{equazione_1}
\end{equation}
where \(\rho\) and \(c\) are, respectively, the density and speed of sound of the liquid medium. \(d\) is the radius of the cylindrical probe. The term \(\omega\) in Eq. \ref{equazione_1} is the angular frequency of the probe (\(\omega=2 \pi f\)), while \(k\) is the wave number, which is equal to \(k=(2\pi)/\lambda\), where \(\lambda\) is the wavelength of the acoustic wave (\(\lambda=c/f\)). 
For further details, we invite the reader to refer to a previous contribution of the authors \cite{viciconte2023towards}.
\subsection{Estimation of the bubble expansion velocity}
\begin{figure*}[h!]
    \centering
\includegraphics[width=1\linewidth]{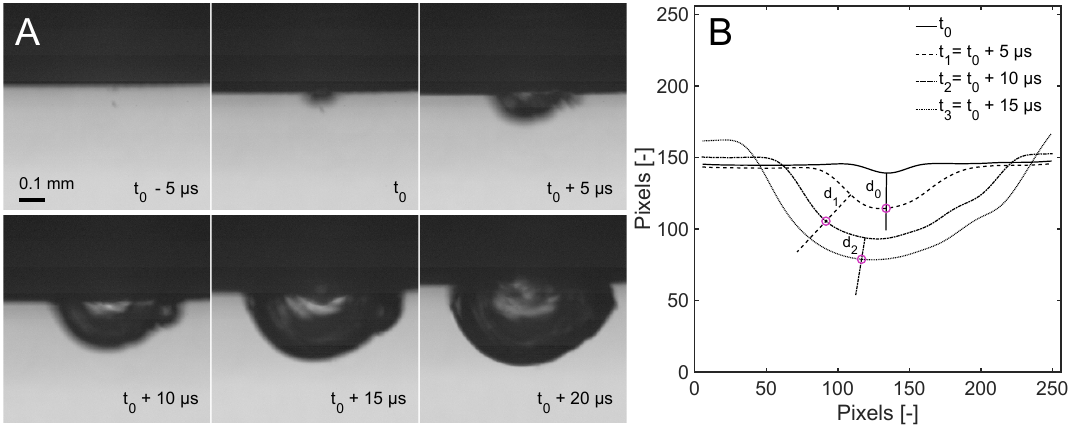}
    \caption{(a) Sequence of frames that capture a bubble's nucleation and expansion in acetone. (b) Illustration of the tracking procedure implemented for tracking the bubble expansion in time.}
    \label{figure3}
\end{figure*}

The high spatial and temporal resolution (respectively  1.74 $\mu$m and 5 $\mu$s), obtained with the optical setup, allowed us to study the behavior of the bubble right after its nucleation. After the onset, the cavitation bubble undergoes rapid expansion. The expansion of the nucleated bubble was tracked starting from the high-speed frames. The sequence shown in Fig. \ref{figure3}a represents the nucleation of a bubble in acetone. Since in most of the experimental run, just a portion of the bubble was visible, a procedure based on the equivalent radius of the half bubble cannot be established. For this reason, to estimate the radial expansion velocity, we proposed a Matlab procedure based on the identification, at every time step, of the vector normal to the bubble boundary. For every frame (Fig. \ref{figure3}a), the bubble boundary was identified by using the function \textit{find} on Matlab. The faction threshold was chosen to preserve the boundary visible in the original grayscale image. Afterward, the data were smoothed using the function \textit{smoothn}, proposed by \cite{Garcia_code,Garcia_2010}. The results for the different frames are visible in Fig. \ref{figure3}b. In a certain location, along the bubble boundary, it is possible to define the normal vector.
At the \(n\)-th time (\(t_n\)), the radial distance traveled by the bubble (\(d_n\)) is defined by the intercept of the normal vector with the boundary at time \(t_{n+1}\) (purple circles in Fig. \ref{figure3}b). Following this notation, the expansion velocity (\(v_n\)) can be computed as: \(v_n=d_n/(t_{n+1}-t_n)\). This procedure allowed us to track the evolution of the bubble over time. Furthermore, it can be applied even in cases where just a portion of the bubble is visible in the frame (Fig. \ref{figure2}a).
To consider the irregular shape of the nucleated bubble, the procedure for estimating the expansion velocity was repeated, considering three vectors normal to the bubble boundary for every time step.

\section{Results}
\begin{figure*}[h!]
    \centering
\includegraphics[width=0.65\linewidth]{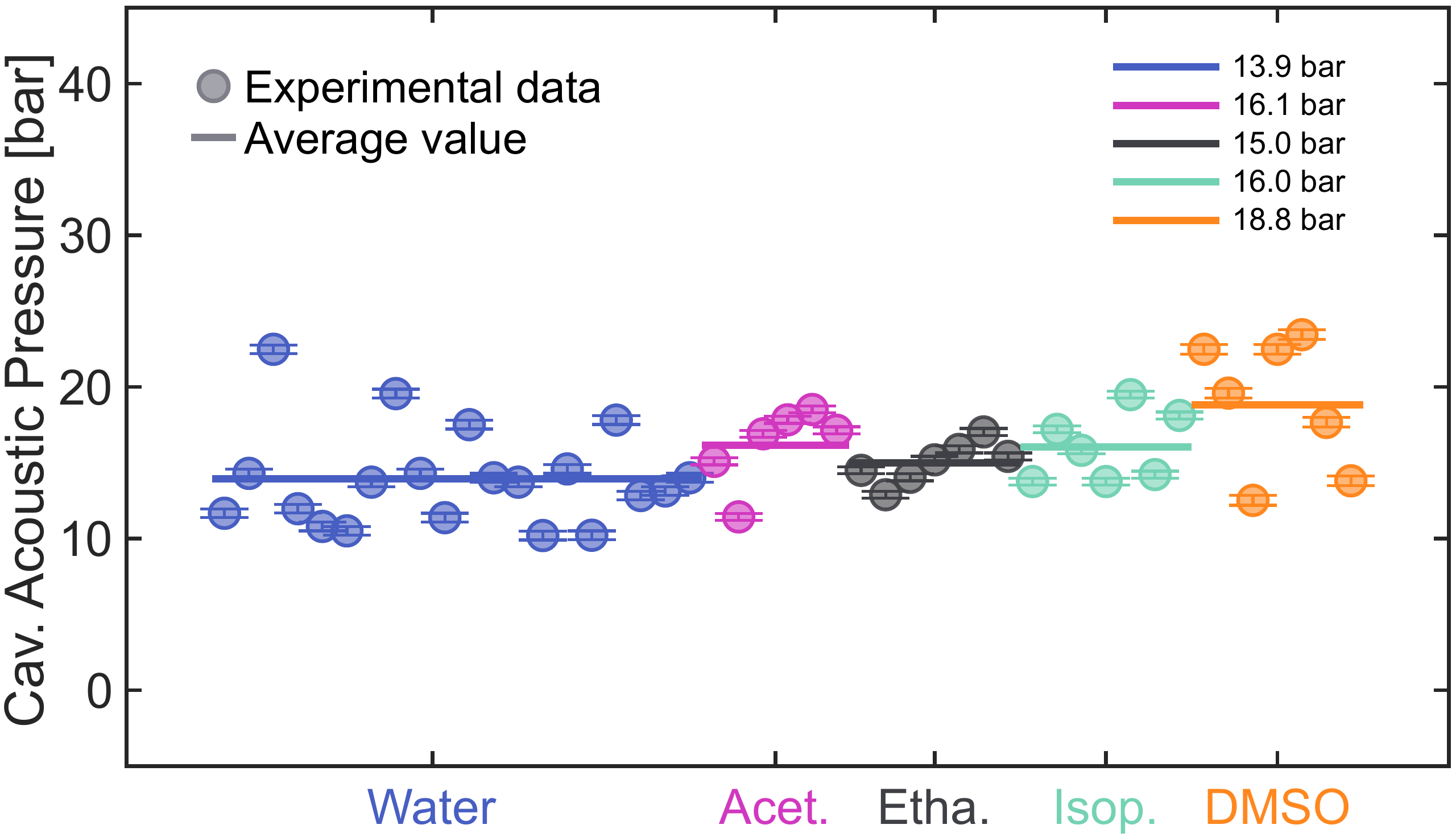}
    \caption{Cavitation acoustic pressure for different liquids tested. The markers indicate the value obtained for every experimental observation, while the solid horizontal lines are the average for every liquid medium. The error bar for every observation is dictated by the optical resolution obtained for the high-speed acquisitions (1.74 $\mu$m/pixel).}
    \label{figure_2}
\end{figure*}

Following the method described in Section \ref{materialsAndMethods}, we measured the acoustic pressure at which cavitation occurs for five different substances. For every substance, the experiment was repeated at least six times. The results are summarized in the scatter plot in Fig. \ref{figure_2}. 
The water data in Fig. \ref{figure_2} are related to three different types tested (\textit{section} 2.1), which have exhibited similar behavior in terms of the average value of cavitation acoustic pressure ( \(p_{0}|_{cav}\)) of 13.6 bar for Milli-Q water, 15.1 bar for water extra pure deionized, and 13.3 bar for pure demineralized water. 
Unexpectedly, the value of the cavitation threshold was similar for the different liquids. Only DMSO was slightly higher than the others. Another important observation was that even the deviation of values around the average threshold was distributed similarly for all fluids.
The behavior exhibited is counterintuitive, considering that the liquid media are characterized by different bulk physical properties, like surface tension ($\sigma$), saturated vapor pressure ($p_V$), and dynamic viscosity ($\mu$L) (Tab. \ref{tab1}) \cite{kim2016pubchem,linstorm1998nist,kaatze1990ultrasonic}.

\begin{figure*}[h!]
    \centering
\includegraphics[width=1\linewidth]{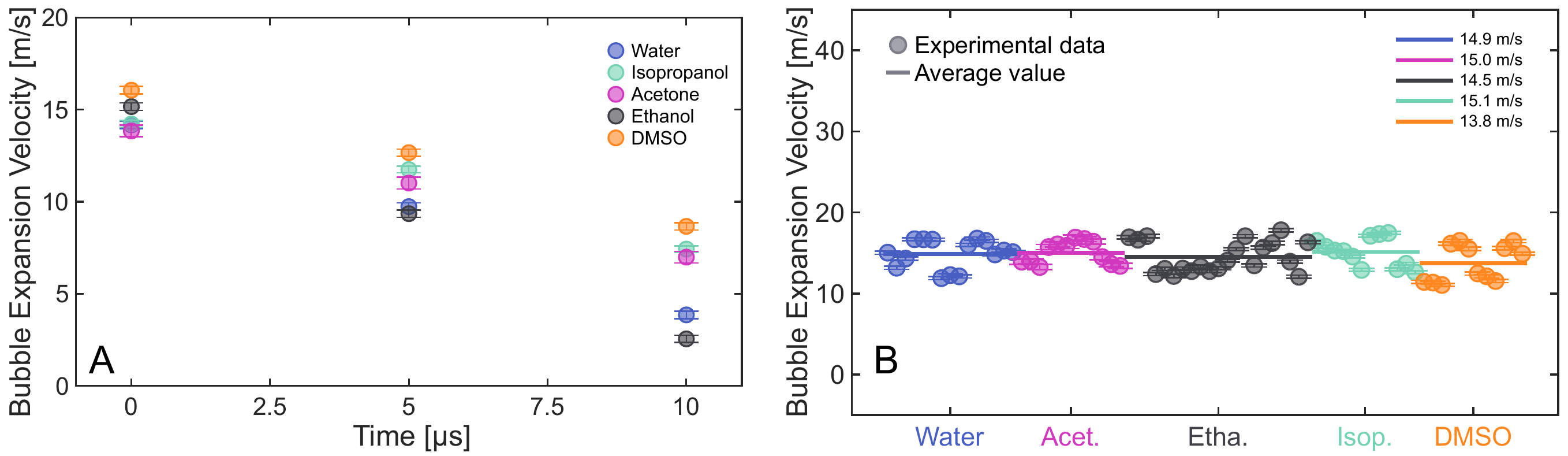}
    \caption{(a) Radial expansion velocity over time computed for one experimental case for every liquid. (b) Expansion velocity (\(v_0\) in Fig. \ref{figure_2}) of the nucleated bubble for the different substances. All bubbles expand at a rate of 10 and 20 m/s. For every experimental observation, \(v_0\) was computed three times, considering three different vectors normal to the bubble boundary (\(d_0\) in Fig. \ref{figure_2}).}
    \label{figure_3}
\end{figure*}

The cavitation threshold does not depend substantially on the liquid medium's surface tension and other properties. The latter is in contradiction with the prediction of the CNT, where the cavitation threshold scales with the surface tension to the power of 3/2 \cite{herbert2006cavitation, brennen2014cavitation, blander1979bubble, skripov1974metastable, caupin2012exploring}.
The impact of the fluid properties on nucleation can also be studied by looking at the dynamics of the cavity right after its onset. By the high spatial and temporal resolutions obtained in the high-speed imaging (Fig. \ref{figure3}), the evolution of the nucleated cavity was observed in many experimental runs. The bubble expansion velocity was computed for the different liquids using the method described in \textit{section 2.3}. The chart in Fig. \ref{figure_3}a shows the expansion velocity over time for one experimental case for every liquid. As expected, the bubble velocity is at its maximum during the rarefaction phase and tends to decrease in time due to the compression phase of the acoustic wave. 
The scatter chart in Fig. \ref{figure_3}b depicts the bubble expansion velocity related to the first time step (\(v_0\) in Fig. \ref{figure3}) for all the experimental observations.
Also, in this case, the dispersion of the data and the average values are comparable for the different media. 
The similar behavior exhibited by the liquids suggests that bubble nucleation and the subsequent dynamics are not functions of the bulk physical properties (Tab. \ref{tab1}).

The mentioned results could be because in a real laboratory system, where a perfectly continuous liquid is impossible to obtain, the nucleation occurs following a heterogeneous mechanism that does not include the rupture of the liquid but where the cavity forms from impurities or sub-micrometric gas pockets, stabilized on a solid surface \cite{atchley1989crevice, borkent2009nucleation, brennen2014cavitation,lohse2016homogeneous, gao2021effects}.
The following sections provide additional experimental data before presenting further arguments to confirm the conjecture about the nucleation mechanism. These concerns investigate the influence of surface wettability and high viscosity on cavity nucleation.

\begin{table}[h]
\caption{Data and physical properties related to the different liquids tested.}\label{tab1}%
\begin{tabular}{@{}llllllll@{}}
\toprule
Liquid &$(p_{0}|_{cav}^{av}$) [bar]&$(\theta$)&$(\mu_L$) [cP]& $(\sigma$) [mN/m]& $c_{\ce{O2}}$ [mg/L] & $c_{\ce{N2}}$ [mg/L] & $c_{air}$ [mg/L]\\
\midrule
Water&13.9&114$^{\circ}$&0.9&72.6& 8.67 \cite{tokunaga1975solubilities}&13.50 \cite{tokunaga1975solubilities}&22.17\\
Ethanol&15.0&154$^{\circ}$&1.07&21.9&66.03 \cite{schnabel2005henry}&135.74 \cite{schnabel2005henry}&201.77\\
Acetone&16.1&148$^{\circ}$&0.32&23.7&64.33 \cite{windmann2012vapor}&162.02 \cite{windmann2012vapor}&226.35\\
DMSO&18.8&144$^{\circ}$&2.47&43.5&25.78 \cite{franco1990photochemical}&-&-\\
Isopropanol&16.0&150$^{\circ}$&2.03&20.9&68.66 \cite{tokunaga1975solubilities}&133.39 \cite{katayama1976solubilities}&202.05\\
\end{tabular}
\end{table}

\subsection{Influence of the surface wettability}

\begin{figure*}[h!]
    \centering
\includegraphics[width=0.65\linewidth]{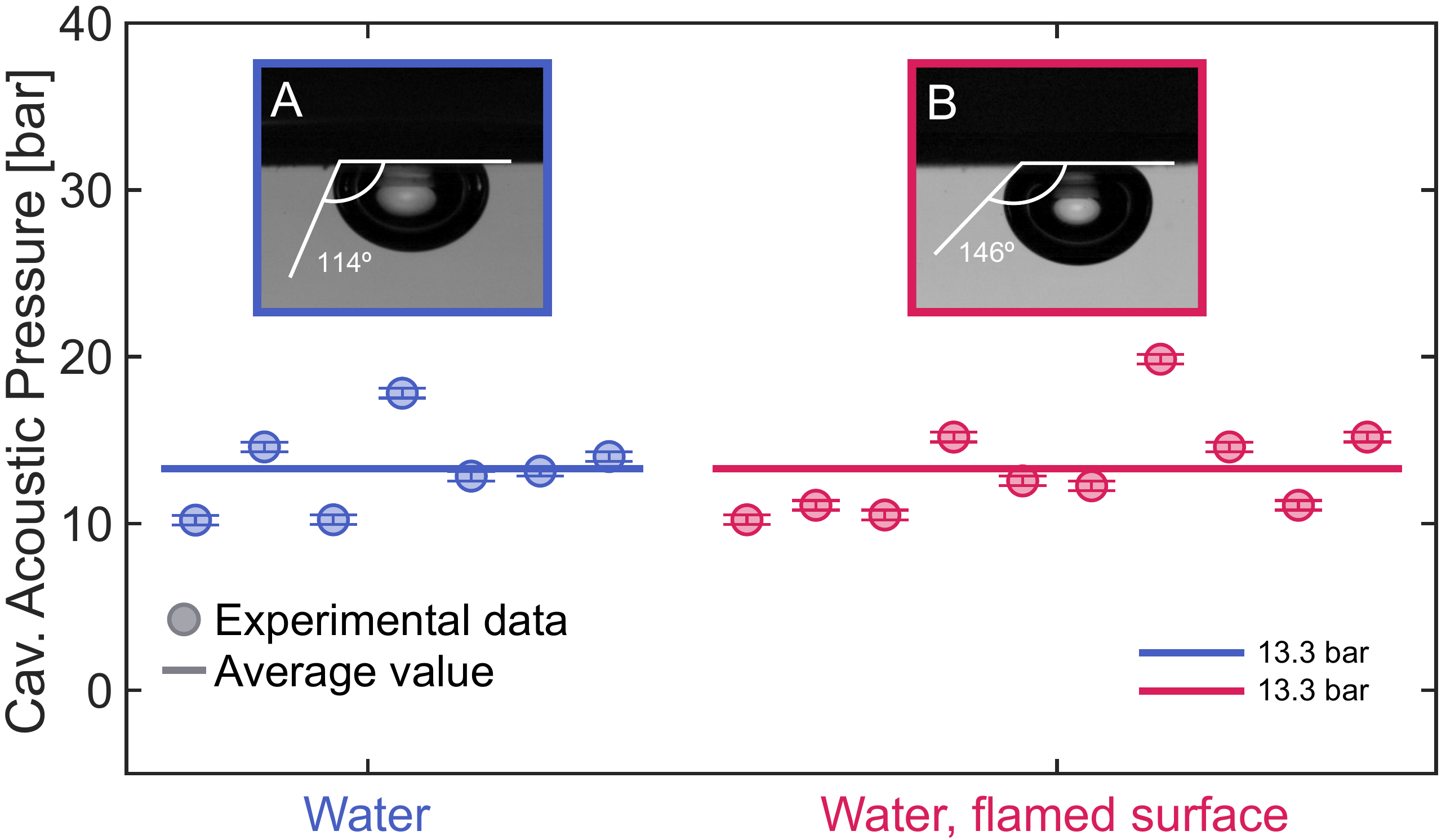}
    \caption{Cavitation acoustic pressure for two different wettability configurations with pure water (demineralized Thermo Fisher). The blue markers indicate the configuration (a): Contact angle measured for an air bubble standing on the untreated surface of the titanium probe. The red markers concern the configuration (b): Contact angle measured for an air bubble standing on the flamed surface of the titanium probe. The level of hydrophilicity substantially increases concerning the case with the untreated surface.}
    \label{figure_4}
\end{figure*}

The nucleation of the first bubble, in the experimental configuration adopted, always happens on the emitting surface of the probe (Fig. \ref{figure2}). Therefore, the surface energy at the interface (wettability) may play an important role in promoting or suppressing the cavity formation. The wettability can be characterized by measuring the contact angle at the interface ($\theta$) \cite{annamalai2016surface}. The contact angle was measured by injecting an air bubble (volume of 0.05 ml) at the interface between the titanium surface of the probe and the liquid medium (Fig. \ref{figure_4}a, b). The values obtained by averaging six different measurements are listed in Tab. \ref{tab1}. There is no marked difference between the contact angle of the different substances, except for that of water, which is markedly lower (114$^{\circ}$ in Tab. \ref{tab1}).

To evaluate whether the surface wettability affects the cavitation threshold, an ideal experimental configuration would be to have two different contact angles with the same liquid. This can be realized for the test with water. By flaming the surface of the titanium probe, before immersing it in the liquid, it is possible to increase the contact angle (Fig. \ref{figure_4}b) and consequently the degree of hydrophilicity, for the case with the untreated surface (Fig. \ref{figure_4}a). This is due to the vaporization of low molecular weight contaminants and impurities on the surface and to the consequent improvement of the adhesion of the surface with the liquid. 
Cavitation experiments have been conducted with both configurations, using pure water demineralized Thermo Fisher (\textit{Water} vs \textit{Water, flamed surface} in Fig. \ref{figure_4}). 
The two contact angles exhibited similar behaviour; indeed, the average value of $p0|_{cav}$ was 13.3 bar (solid lines in Fig. \ref{figure_4}).
This data suggests that in the limited investigated range, the contact angle and the level of hydrophilicity do not affect the onset of acoustic cavitation. 

\subsection{Influence of the liquid viscosity}

Silicon oils offer the possibility to study the cavitation threshold and the nucleation dynamics over a wide range of viscosities while maintaining the other physical properties. The influence of dynamic viscosity on the inception of cavitation was studied by testing three different silicon oils: 5, 10, and 50 cP.
Cavitation amplitude and bubble expansion velocity were estimated through the same experimental procedure adopted for the other substances (Section \ref{materialsAndMethods}).

\begin{figure*}[h!]
    \centering
    \includegraphics[width=1\linewidth]{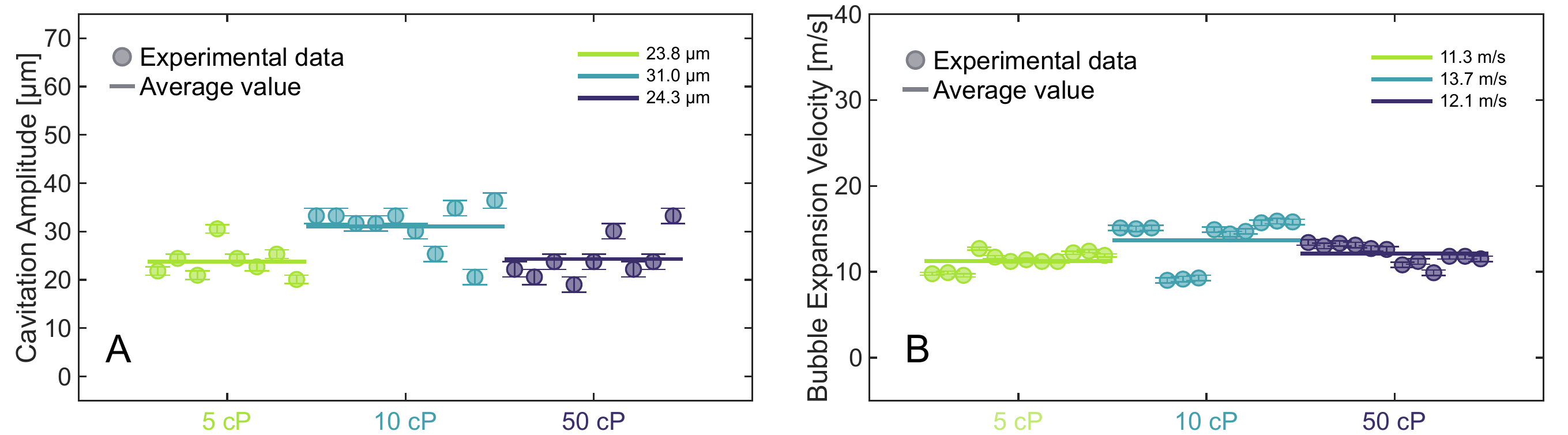}
    \caption{Experimental estimations for silicon oils characterized by three different viscosities: 5, 10, and 50 cP. (a) Cavitation acoustic pressure. (b) Expansion velocity (\(v_0\) in Fig. \ref{figure3}) of the nucleated bubble.}
    \label{figure_5}
\end{figure*}

The experimental data (Fig. \ref{figure_5}) shows that the cavitation amplitude and bubble expansion are comparable for the three different oils. These observations reflect the behavior of the other substances (Fig. \ref{figure_2}) and suggest that the kinematic viscosity does not significantly affect the cavitation threshold and the nucleation dynamics. 

\subsection{Nucleation mechanism and bubble dynamics}

The experimental findings show that the cavitation threshold for 24 kHz acoustic waves is almost constant and does not inherently depend on surface tension and other fluid properties. This is apparently in contradiction with the prediction of the CNT, where the cavitation threshold scales with the surface tension to the power of 3/2 \cite{herbert2006cavitation, brennen2014cavitation, blander1979bubble, skripov1974metastable, balibar2002metastable}. 
This discrepancy is probably attributable to the nucleation mechanism. While the CNT describes a homogeneous nucleation mechanism, where cavitation occurs due to the breaking of molecular bonds in an ideal continuum, in the present laboratory case, cavitation is likely to happen starting from impurities or gas nuclei stabilized on the emitting surface of the probe. Some reference corroborates this hypothesis works \cite{atchley1989crevice, borkent2009nucleation,lohse2016homogeneous}, where the authors estimated the cavitation threshold of nucleation triggered by bubbles trapped in cylindrical sub-micrometric pits, artificially created to act as nucleating sites, applying a single pressure pulse.
The present estimates of the cavitation threshold correspond to the value obtained by the authors \cite{borkent2009nucleation} for gas nuclei stabilized in cavities in the order of 100 nm.
Such a cavity size is compatible with the present experimental conditions (Section 2.2). In fact, despite the tip being cleaned from macroscopic bubbles and impurities before every run, the optical resolution (1.74 $\mu$m/pixel) did not allow the detection of the presence of sub-micrometric gas pockets on the emitting surface of the probe to be excluded.
A sort of pre-seeded nucleation mechanism, where cavitation happens due to the sudden expansion of pre-existing gas pockets, also explains the experimental results for the bubble expansion velocity (Fig. \ref{figure_4}). The expansion of gas nuclei is a fast and dynamic process where the bulk physical properties of the liquid media do not play a relevant role.
The phenomenon can also be analyzed by solving the well-known Rayleigh–Plesset equation (RP), which for an isothermal case takes the following form:

\begin{equation}
\frac{p_{V}-p_{\infty(t)}}{\rho_L} + \frac{p_{G_{0}}}{\rho_L}\left(\frac{R_0}{R}\right)^3=R\frac{d^2R}{dt^2}+\frac{3}{2} \left(\frac{dR}{dt} \right)^2+\frac{4\nu_L}{R}\frac{dR}{R}+\frac{2\sigma}{\rho_LR},
\label{equazione_2}
\end{equation}
where $t$ is the time variable, $p_{\infty(t)}$ is the driving pressure, and $p_{G_{0}}$ is the partial pressure of the initial gas contained in the bubble. $R$ and $R_0$ are the bubble and initial radius, respectively. The properties of the liquid appearing in the equation are: vapor pressure ($p_{V}$), density ($\rho_L$), and kinematic viscosity ($\nu_L$).

\begin{figure*}[h!]
    \centering
    \includegraphics[width=1\linewidth]{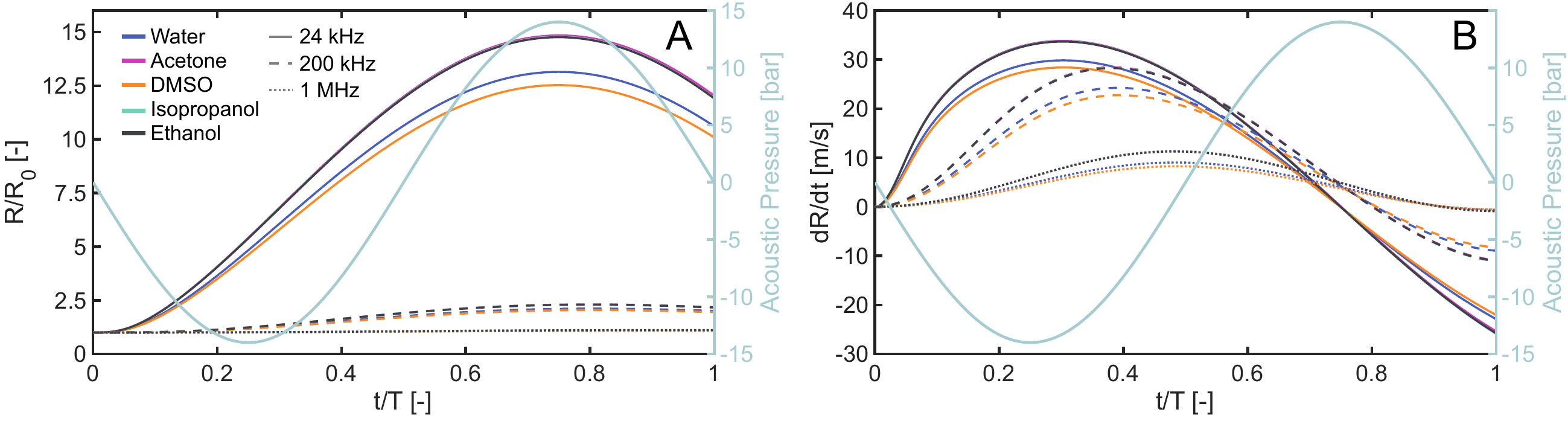}
    \caption{Numerical solutions of the RP. (a) Time evolution of the bubble radius for different liquids and three pressure wave frequencies. (b) Expansion velocity of the bubble radius for different liquids and three pressure wave frequencies. The curves related to isopropanol and acetone overlap with those of ethanol since these three solvents present an identical behavior.}
    \label{figure_6}
\end{figure*}

The RP is derived from the Navier–Stokes equations (NS), enforcing a circular symmetry to describe the dynamics of a spherical bubble. For this reason, the physical validity of the RP stands on the same hypothesis behind the NS. The fundamental hypothesis, for the validity of the NS, is that of a continuum medium. The continuum theory is valid when the length scale is larger than the mean free path, which can be considered, for a liquid medium, in the same order of magnitude of the distance between the molecules \cite{graziano2004water}. This implies that the NS preserve their validity for liquids down to a length in the order of a few nanometers. The RP can be used to model the expansion of a bubble with an initial diameter in the sub-micrometric scale. This statement is demonstrated in literature by two different frameworks: the first one where the RP solution was validated with Molecular Dynamics simulations for the expansion of nanometric bubbles \cite{man2018rayleigh, tsuda2015validation, holyst2010large, okumura2003nonequilibrium}. In the second one, Gallo et al. \cite{gallo2020nucleation} proposed a novel model based on an isothermal diffuse interface description of a two-phase liquid–vapor system endowed with thermal fluctuations, exploiting Landau and Lifshitz’s fluctuating hydrodynamic theory. The results, obtained by the authors, suggest that the Tolman correction and the diffuse interface are crucial to describe the nucleation process. However, following the nucleation phase, the model prediction in the mesoscale agrees with the RP solution.
In the present case, to study the expansion of a gas nucleus, the RP was solved numerically, assuming an initial radius ($R_0$) of 100 nm, a wave frequency of 24 kHz, and an acoustic pressure amplitude of 14 bar. Fig. \ref{figure_6} shows the results obtained in non-dimensional terms.

The temporal evolution of the bubble radius is similar between the different liquids, especially during the first phase of rapid expansion (solid lines in Fig. \ref{figure_6}b).
The time derivative of the radius also exhibits an almost identical trend for the various liquids (solid lines in Fig. \ref{figure_6}b).
The curves related to isopropanol and acetone overlap with those of ethanol 
due to their almost identical physical properties.
The similarity of the solution for the different liquids is to be attributed to the higher-order inertial term ($R\frac{d^2R}{dt^2}$), present in the RP equation (Eq. \ref{equazione_2}), which is dominant compared to the terms containing the liquid properties.
The RP numerical solutions confirm what has been observed experimentally by measuring the expansion velocity of the bubble (Fig. \ref{figure_4}).
To better understand the phenomenology behind the bubble expansion, the RP equation was also solved for other forcing frequencies by keeping the same acoustic pressure amplitude of the 24 kHz case (14 bar). The numerical solution and its derivative for the cases related to 200 kHz and 2 MHz are shown in Fig. \ref{figure_6}.
It is compelling to observe that the formation of a macroscopic bubble is suppressed when the frequency of the acoustic wave increases.

The growth suppression, occurring at high frequencies, results from an acoustic wave period smaller than the nucleus expansion's characteristic time.
In the context of heterogeneous nucleation, this mechanism could explain the increase of the cavitation threshold with the wave frequency \cite{viciconte2023towards, urick1975principles}.
In summary, the experimental and numerical data suggest that heterogeneous cavitation is not inherently specific to the fluid but primarily depends on the perturbation rate.
Another quantity that could potentially affect the nucleation process is the concentration of dissolved air $c_{air}$, that can be expressed as a summation of the concentrations of dissolved oxygen and nitrogen (respectively $c_{\ce{O2}}$ and $c_{\ce{N2}}$). $c_{\ce{O2}}$ and $c_{\ce{N2}}$ were estimated through the Henry’s law constants available in the literature \cite{tokunaga1975solubilities, schnabel2005henry, windmann2012vapor, franco1990photochemical, katayama1976solubilities}, and by taking into account the experimental conditions (saturation level at 1 atm). The results are listed in Tab. \ref{tab1}. The dissolved air concentration is one order of magnitude higher in the solvents compared to water. Since this difference is not reflected in the values of the cavitation threshold for the corresponding substances, the concentration of dissolved air is unlikely to affect the nucleation. This suggests that heterogeneous nucleation is promoted by the gas nuclei present in the medium and not by the amount of dissolved air.

\section{Conclusion}\label{sec13}

The experimental findings presented in this study indicate that the occurrence of cavitation is influenced by the rate at which the fluid transitions to its metastable state. Ultrasonically induced cavitation experiments were performed on various liquids using a horn-type ultrasonic device operating at 24 kHz. The results demonstrate that the cavitation threshold is largely unaffected by variations in the fluid's bulk properties and thermodynamic characteristics. Since bulk properties are evaluated under equilibrium conditions, we propose that nucleation cannot be adequately described as a sequential progression through equilibrium states.
To further explore this hypothesis, we investigated, through high-speed imaging, the expansion of the cavitation bubble in the first time instant after its nucleation. The same phenomenon was also studied by solving the Rayleigh–Plesset equation. Both experimental and numerical results confirmed that, under rapid perturbations, differences in fluid responses become negligible. These observations suggest that, in heterogeneous nucleation, the cavitation threshold is not inherently determined by the intrinsic properties of the fluid but is primarily governed by the perturbation rate. Specifically, the solution of the Rayleigh–Plesset also indicates that liquids can withstand higher pressure differentials when the frequency of the forcing pressure wave is higher.


\bibliographystyle{unsrt}  
\bibliography{references}  

\end{document}